\documentclass[prb,twocolumn,showpacs,preprintnumbers,amsmath,amssymb]{revtex4}
\usepackage{graphicx}
\usepackage{dcolumn}
\usepackage{bm}
\usepackage{color}

\newcommand{\note}[1]{}

\newcommand{\Tc}{\ensuremath{T_{\mathrm{C}}}}

\begin{document}

\title{First-principles prediction of high Curie temperature for ferromagnetic 
bcc-Co and bcc-FeCo alloys and its relevance to tunneling magnetoresistance}

\author{M.~Le\v{z}ai\'{c}}\email{M.Lezaic@fz-juelich.de}
\author{Ph.~Mavropoulos}\email{Ph.Mavropoulos@fz-juelich.de}
\author{S.~Bl\"ugel}

\affiliation{Institut f\"ur Festk\"orperforschung, Forschungszentrum
  J\"ulich, D-52425 J\"ulich, Germany}

\date{\today}

\begin{abstract}
We determine from first-principles the Curie temperature \Tc\ for bulk
Co in the hcp, fcc, bcc, and tetragonalized bct phases, for FeCo
alloys, and for bcc and bct Fe. For bcc-Co, \Tc=1420~K is
predicted. This would be the highest Curie temperature among the Co
phases, suggesting that bcc-Co/MgO/bcc-Co tunnel junctions offer high
magnetoresistance ratios even at room temperature.  The Curie
temperatures are calculated by mapping {\it ab initio} results to a
Heisenberg model, which is solved by a Monte Carlo method.
\end{abstract}


\maketitle

In the past few years we are witnessing a compelling race of different
research
groups\cite{Parkin:04.1,Yuasa:04.1,Djayaprawira:05.1,Ikeda:05.1}
hunting the maximum tunneling magneto-resistance ratio (TMR) of
magnetic tunnel junctions\cite{Tsymbal03} (MTJs) at room temperature
(RT). MTJs, made of two ferromagnetic electrodes separated by an
insulating barrier, open vistas to a wide field of technological
applications, in particular in non-volatile magnetic random access
memory, or new type of recording heads for ultrahigh-density hard-disk
drives. Large temperature stability and other attributes are required
which can be met by a high TMR at RT.

After the first reported observations of TMR at room
temperature\cite{Miyazaki:55.1} in Al$_2$O$_3$-based junctions,
reaching 30\% at 4.2~K and 18\% at 300~K, the achievement of high TMR
values developed into a challenging materials science problem.  Guided
by theoretical predictions of extremely high TMR over 1000\% for
Fe/MgO/Fe\cite{Mathon:01.1}, giant values of over 150\% at RT have been
experimentally achieved for fully epitaxial
Fe(100)/MgO(100)/Fe(100)\cite{Yuasa:04.1} and
Fe(100)/MgO/Co(100),\cite{Yuasa:05.1} as well as poly-crystalline
FeCo/MgO/FeCo \cite{Parkin:04.1} and FeCoB/MgO/FeCoB(001)
MTJs.\cite{Djayaprawira:05.1} Theory predicts even higher TMR values
(at temperature $T=0$) for epitaxial Co/MgO/Co
junctions\cite{Zhang:05.1} than for Fe/MgO/Fe.  Bcc-Co is a metastable
phase and cannot be grown as a single crystal. Recently, however, it
has been possible to grow bcc Co in contact with
MgO,\cite{Schneider06,Yuasa:05.1,Yuasa:06.1} and epitaxial
bcc-Co/MgO/bcc-Co(100) MTJs show a high TMR, unusually
stable\cite{Yuasa:06.1} with $T$.

In most MTJs there is a large difference of the TMR between cryogenic
and room temperature. The TMR, {\it i.e.}, the relative change of
resistance when the two ferromagnetic leads are coupled either
ferromagnetically or antiferromagnetically, depends on the details of
the electronic structure, such as the spin polarization of the
ferromagnet at the Fermi level $E_F$,\cite{Julliere:75.1} or the
spin-dependent symmetry of the states at
$E_F$.\cite{Mavropoulos00,Mathon:01.1} At $T>0$, magnetic excitations
mix the two spin channels, decreasing the TMR.\cite{Itoh00} One can
infer that a high \Tc\ and spin stiffness entail a temperature
stability of the magnetic structure and of the TMR. In this paper we
investigate the Curie temperature of bcc-Co and Fe$_{1-x}$Co$_x$
alloys and relate it to fcc-Co, hcp-Co, and bcc-Fe, concluding that a
high (calculated)\cite{footnoteb} \Tc\ of bcc-Co is responsible for
the temperature stability\cite{Yuasa:06.1} of TMR in bcc-Co/MgO/bcc-Co
junctions.

We calculate the Curie temperature using a standard recipe: the
adiabatic approximation for the calculation of magnon
spectra.\cite{Liechtenstein87,Halilov98} {\em Ab initio} total-energy
results are mapped to the classical Heisenberg model,
\begin{equation}
H=-\sum_{i,j\,;\,i\neq j}J_{ij}\,\vec{e}_i\cdot \vec{e}_j.
\label{eq:1}
\end{equation}
Here, $J_{ij}$ are the exchange constants between the magnetic
moments at sites $i$ and $j$, and $\vec{e}_i$ is a unit vector along
the  moment of atom $i$. \Tc\ was calculated within this model
by a Monte Carlo method (using 1728 atoms in the supercell) by
locating the susceptibility peak.

The {\em ab initio} results are calculated within the generalized
gradient approximation\cite{GGA} to density-functional theory. We
employ the full-potential Korringa-Kohn-Rostoker (KKR) Green function
method\cite{SPRTBKKR}, using the coherent potential approximation for
the electronic structure of the Fe$_{1-x}$Co$_x$ random alloys. The
exchange constants $J_{ij}$ are related to the Green function via the
Lichtenstein formula\cite{Liechtenstein87} within perturbation theory
assuming an infinitesimal direction change of the magnetic moments at
sites $i$ and $j$. For Fe and Co we also compare to a complementary
method, where finite-angle static magnon dispersion relations,
$J(\vec{q})$, are calculated on a dense mesh of $\vec{q}$-points in
the Brillouin zone within the full-potential linearized augmented
plane-wave method (FLAPW).\cite{FLEUR} From $J(\vec{q})$, the
real-space constants $J_{ij}$ are found via a Fourier
transform.\cite{Halilov98}  The two methods give the same trends. We
stress that the trends are important for our conclusions, and not
the absolute values of \Tc, which are off the experimental value (bcc
Fe: $\Tc=1043$~K; fcc Co: $\Tc=1403$~K)\cite{Bozorth51,footnoteb} by
$\sim$10\%.

\begin{figure}
\includegraphics[width=\linewidth]{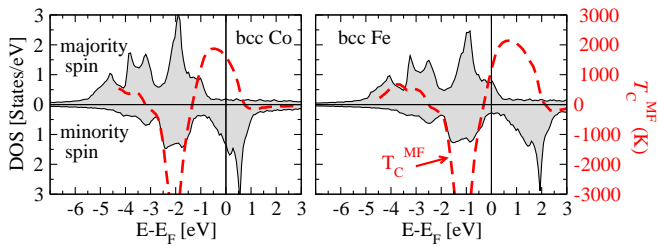}
\caption{(color online) Density of states (solid line) and mean-field ordering
   temperature (dashed line) for bcc Co (left) and bcc Fe
   (right). \label{fig:dos+tcmf}}
\end{figure}

Our calculations include bcc Fe and Co, tetragonalized (bct) Fe and Co
(considering growth on MgO(001)), the disordered alloys
Fe$_{0.75}$Co$_{0.25}$, Fe$_{0.50}$Co$_{0.50}$, and
Fe$_{0.75}$Co$_{0.25}$, and the ordered alloy FeCo (in the CsCl
structure). In order to see the effect of compression on the magnetic
properties, we calculated the results at two lattice parameters:
a=2.30~au=2.804~\AA, corresponding to the calculated equilibrium
lattice parameter of bcc Co, and a=2.40~au=2.857~\AA.  We find for hcp
Co an equilibrium lattice parameter ${\rm a}=2.488$~\AA, with ${\rm
c/a}=1.623$ (experimental values are ${\rm a}=2.51$~\AA, ${\rm
c/a}=1.623$), for fcc Co ${\rm a}=3.52$~\AA\ (exp.\ value is ${\rm
a}=3.54$~\AA), for bcc Co $a=2.804$~\AA\ (extrapolated exp.\ value is
${\rm a}=2.819$~\AA),\cite{Ellis41} and for bcc Fe a=2.825~\AA\ (exp.\
value is a=2.867\AA). For Co and Fe on MgO(001) we find a mismatch of
5\% in the surface lattice parameter (compared to
$\sqrt{2}\mathrm{a}_{\rm MgO}=5.63$~\AA), resulting in a ratio
c/a=0.857 for bct Co and c/a=0.909 for bct Fe. The ground state spin
moments per Co atom, $M^{\rm hcp}=1.6$, $M^{\rm fcc}=1.65$, and
$M^{\rm bcc}=1.75\ \mu_B$, change with the crystal structure by less
than 10\%.  The density of states (DOS) of bcc Co, shown in
Fig.~\ref{fig:dos+tcmf}, exhibits the typical bimodal behavior of the
bcc lattice with bonding and antibonding states. The spin polarization
at $E_F$, $P=-82$\%, is negative and has a larger (absolute) value
than for bcc-Fe ($P=52$\%). Our calculated band structure agrees with
the one of [\onlinecite{Callaway}].

\begin{table*}
\begin{tabular}{ccccccccccc}
\hline
\hline
Alloy & a(\AA) (c/a) & $M_{\mathrm{Fe}}$ & $M_{\mathrm{Co}}$ &
$\overline{M}$ & $\overline{M}_{\mathrm{exp}}$[\onlinecite{Bardos69}] &
$J_1^{\mathrm{Fe-Fe}}$ &
$J_1^{\mathrm{Co-Co}}$ & $J_1^{\mathrm{Fe-Co}}$ &$\Tc^{\mathrm{KKR}}$  & $\Tc^{\mathrm{FLAPW}}$ \\
\hline
Fe bcc & 2.804 & 2.107 & $-$ & 2.107 & 2.22 & 17.45 & $-$ & $-$ & 970 & 1120 \\
   & 2.857 & 2.197 & $-$ & 2.197 &      & 15.25 & $-$ & $-$ & 900 &\\
Fe bct & 2.978 (0.909) & 2.321 & $-$ & 2.321 & $-$ & 12.71 & $-$ & $-$ & 750 &\\
\hline
Fe$_{0.75}$Co$_{0.25}$ & 2.804 & 2.421 & 1.785 & 2.263 & 2.45 & 25.80 & 24.54 & 31.78 & 1390 &\\
 (disordered bcc)          & 2.857 & 2.512 & 1.803 & 2.335 &      & 26.93 & 24.61 & 31.67 & 1490 &\\
\hline
Fe$_{0.50}$Co$_{0.50}$ & 2.804 & 2.523 & 1.778 & 2.151 & 2.35 & 28.55 & 22.05 & 30.76 & 1600 &  \\
 (disordered bcc)          & 2.857 & 2.588 & 1.801 & 2.194 &      & 27.99 & 22.13 & 30.27 & 1600 &\\
\hline
FeCo                   & 2.804 & 2.729 & 1.727 & 2.228 & 2.42 &  $-$  &  $-$  & 28.84 & 1660 &\\
(ordered CsCl struct.) & 2.857 & 2.803 & 1.743 & 2.273 &      &  $-$  &  $-$  & 28.50 & 1670 &\\
\hline
Fe$_{0.25}$Co$_{0.75}$ & 2.804 & 2.560 & 1.759 & 1.959 & $-$  & 26.64 & 20.01 & 28.33 & 1520 & \\
 (disordered bcc)          & 2.857 & 2.633 & 1.791 & 2.002 &      & 26.42 & 20.06 & 28.01 & 1540 &\\
\hline
Co bcc & 2.804 & $-$ & 1.751 & 1.751 & $-$ & $-$ & 18.48 & $-$ & 1420 & 1670\\
   & 2.857 & $-$ & 1.790 & 1.790 &     & $-$ & 18.71 & $-$ & 1370 &\\
Co bct & 2.978 (0.857)& $-$ & 1.740 & 1.740 & $-$ & $-$ & 16.76 & $-$ & 1380 &\\
Co fcc & 3.519        & $-$ & 1.646 & 1.646 & $-$ & $-$ & 13.82 & $-$ & 1280 & 1200\\
Co hcp & 2.487 (1.623)& $-$ & 1.602 & 1.602 & $-$ & $-$ & 14.66 & $-$ & 1300 & 1350\\
\hline
\hline
\end{tabular}
\caption{Calculated (within KKR) magnetic moments (in
  $\mu_B$), first-neighbor exchange constants $J_1$ (in meV), and
  \Tc (in K) of bcc FeCo alloys, bcc Fe and Co, bct Fe and Co,
  fcc and hcp Co. The bcc structures are considered at two lattice
  constants, to show the effect of compression. For the bct
  structures, the in-plane lattice constant was adapted to MgO and c/a
  was relaxed. \Tc\ is calculated by a Monte Carlo method with exchange constants derived
  from the KKR or FLAPW methods.\label{tab:1}}
\end{table*}

Our results on the magnetic properties are summarized in
Table~\ref{tab:1}. A striking effect is that the exchange constants
and the \Tc\ increase with the Co concentration in Fe$_{1-x}$Co$_x$
alloys, with a maximum of about \Tc=1670~K for the ordered FeCo alloy,
and a $\Tc\approx1400$~K for bcc and bct Co. In contrast, bct Fe
suffers a 20\% decrease of \Tc\ compared to bcc Fe. We also find
interesting trends in the magnetic moments and in the
lattice-parameter dependence of the exchange constants.  As the Co
concentration increases, the local moment of Fe becomes larger,
climbing from $M_{\mathrm{Fe}}\approx2.2~\mu_B$ in pure Fe to about
2.6~$\mu_B$ in Fe$_{0.25}$Co$_{0.75}$. The Co moment is comparatively
independent of concentration, about 1.73--1.80~$\mu_B$. Moreover, the
Fe-Fe first-neighbor exchange constants $J_1$(Fe-Fe) are strongly
dependent on the lattice parameter for low Co concentrations $x$, but
much less so for high $x$; the Co-Co and Fe-Co exchange constants,
$J_1$(Co-Co) and $J_1$(Fe-Co), are much less affected by the lattice
parameter at any $x$. The same behavior is found for the
longer-distance exchange constants (not shown here). This leads to a
strong dependence of \Tc\ on the lattice parameter for Fe-rich alloys,
and a weak dependence for Co-rich alloys.

The change of the Fe properties upon increasing the Co concentration
can be explained by observing that Fe progressively changes character
from a weak ferromagnet (i.e., both majority- and minority-spin
$d$-states, $d^{\uparrow}$ and $d^{\downarrow}$, are only partly
occupied for pure Fe) to a strong ferromagnet ($d^{\uparrow}$ states
become fully occupied); Co itself is a strong ferromagnet. The
progressive change occurs because the $d$(Fe)-$d$(Co) hybridization is
weaker than the $d$(Fe)-$d$(Fe) hybridization, as the Co $d$ states
are more localized (they are deeper in the atomic potential well) than
the Fe $d$ states. As a result, an increased coordination with Co
results in an increase of the Fe moment and exchange splitting,
driving the Fe $d^{\uparrow}$ states lower and the Fe $d^{\downarrow}$
states higher in energy.  The Co moment, on the other hand, is rather
stable (around 1.8~$\mu_B$), since the Co $d^{\uparrow}$ states are
always fully occupied.\cite{footnotea} The average moment per unit
cell, $\overline{M}$, first rises upon alloying with Co, due to the
rapid increase of the Fe local moment, then peaks and drops for higher
Co content (because of the comparatively lower Co moment). This peak
of $\overline{M}$ in Fe-Co alloys is long
known\cite{Bardos69,Collins63,KublerBook,Diaz06}, and is related to
the triangular shape of the Slater-Pauling curve at the center of the
$3d$ series\cite{KublerBook}; density-functional calculations
reproduce this behavior, although they underestimate the local Fe
moment by about 10\% in these alloys.\cite{Diaz06}

In a weak ferromagnet the magnetic properties can be easily affected
by structural perturbations (compression or tetragonalization) because
an extended part of the Fermi surface has $d^{\uparrow}$ and
$d^{\downarrow}$ character, allowing for
$d^{\uparrow}$-$d^{\downarrow}$ charge transfer upon the structural
perturbation. Such $d^{\uparrow}$-$d^{\downarrow}$ transfer can affect
the local moment, but even more the spin susceptibility (and exchange
constants $J_{ij}$) which is sensitive to
$d^{\uparrow}$-$d^{\downarrow}$ Fermi-surface crossing or
nesting\cite{Lizarraga04} because of virtual
$d^{\uparrow}$-$d^{\downarrow}$ spin-flip excitations. In the case of
strong ferromagnetism (Co or FeCo alloys), small structural
perturbations cannot cause a $d^{\uparrow}$-$d^{\downarrow}$ charge
transfer, because the $d^{\uparrow}$ band is fully occupied and well
under $E_F$. Thus the exchange interactions are more stable in Co and
FeCo alloys.

We proceed to the discussion of the exchange interactions and \Tc\ by
introducing\cite{Liechtenstein87} the coefficient $J_0=\sum_{i\neq
0}J_{0i}$, corresponding to the band-energy cost for flipping the
magnetic moment of a single atom, reflecting a ``single-site spin
stiffness''. It is related to the mean-field Curie temperature via
$k_B \Tc^{\mathrm{MF}}=2J_0/3$ ($k_B$ is Boltzmann's constant). (It is
well-known that mean-field theory overestimates \Tc, but it is a
useful tool for trends analysis.) By treating $E_F$ as a parameter, we
calculate $\Tc^{\mathrm{MF}}(E)$ as a function of band filling; in
this way we are able to see the individual contribution of the states
at each energy to the exchange interactions.  In
Fig.~\ref{fig:dos+tcmf} we show $\Tc^{\mathrm{MF}}(E)$, together with
the DOS, for bcc Fe and Co. Evidently, $\Tc^{\mathrm{MF}}(E)$ has a
very similar form for Fe and Co, up to the approximately rigid band
shift (it also has slightly higher values for Fe because of the
stronger $d$-$d$ hybridization); thus, Fe and Co can be compared in a
unified picture. The negative values of $\Tc^{\mathrm{MF}}(E)$,
peaking around $-2$~eV for Co and $-1$~eV for Fe, indicate an
antiferromagnetic coupling, known from
$\delta$-Mn.\cite{KublerBook} At higher energies the double-exchange
mechanism sets in and $\Tc^{\mathrm{MF}}(E)$ obtains strong positive
contributions as $E$ crosses the final part of the $d^{\uparrow}$
states and the anti-bonding $d^{\downarrow}$ states. Finally it drops
to zero once the $d^{\downarrow}$ states are filled and the exchange
mechanism is no more present.

In Fe, $E_F$ is located at a steep, ascending point of
$\Tc^{\mathrm{MF}}(E)$, before the maximum. Therefore, small
structural perturbations resulting in band shifts have a strong
influence the \Tc\ of Fe, as seen in Table~\ref{tab:1}. For Co,
$\Tc^{\mathrm{MF}}(E)$ is already descending at $E_F$, but is not as
steep as for Fe. From this argument, Co is expected to have a higher
and more robust \Tc\ than Fe. In a FeCo alloy $E_F$ is in-between, at
the maximum of $\Tc^{\mathrm{MF}}(E)$; then one expects the highest
and most robust \Tc, as is found by the Monte Carlo calculations.

In conclusion, ferromagnetism is found to be more robust in bcc Co
than in bcc Fe. The calculated \Tc(Co) is higher and stable with
respect to structural changes, even in the tetragonalized (bct)
structures. This can be advantageous in the temperature dependent TMR
of Co/MgO/Co MTJs compared to Fe/MgO/Fe, as observed in recent
experiments.\cite{Yuasa:06.1} The \Tc\ of bcc Co is calculated to be
the highest among all Co phases. Although the highest \Tc\ was found
for FeCo alloys, they have the drawback that the TMR can drop if FeCo
disordered layers appear at the interface with MgO, because the
symmetry, enforcing spin polarization of the current, will be absent.
Taking the excellent coherent spin-dependent transport properties in
conjunction with the MgO(100) barrier into consideration, Co/MgO/Co
may be an ideal junction for high TMR values at elevated temperatures.


We gratefully acknowledge enlightening discussions with
S.~Yuasa and P.~H.~Dederichs.


\begin{thebibliography}{99}


\bibitem{Parkin:04.1}
S.S.P.~Parkin, C.~Kaiser, A.~Panchula, P.M.~Rice, B.~Hughes, M.~Samant,
and See-Hun Yang,
Nature Materials {\bf 3}, 862 (2004).

\bibitem{Yuasa:04.1}
S.~Yuasa, T.~Nagahama, A.~Fukushima, Y.~Suzuki, and K.~Ando, 
Nature Materials {\bf 3}, 868 (2004).

\bibitem{Djayaprawira:05.1}
K.~Tsunekawa, D.D.~Djayaprawira, M.~Nagai, H.~Maehara, S.~Yamagata,
N.~Watanabe, S.~Yuasa, Y.~Suzuki, and K.~Ando, 
Appl.~Phys.~Lett.~{\bf 87}, 072503 (2005);
J. Hayakawa, S.~Ikeda,  F.~Matsukura, H.~Takahashi, and H.~Ohno, 
Jpn.~J.~Appl.~Phys.~{\bf 44}, L587 (2005).

\bibitem{Ikeda:05.1}
S.~Ikeda, J.~Hayakawa, Y.M.~Lee, R.~Sasaki, T.~Meguro, F.~Matsukura,
and H.~Ohno, 
Jpn.~J.~Appl.~Phys.~{\bf 44}, L1442 (2005).

\bibitem{Tsymbal03}
E.Y. Tsymbal, O.N. Mryasov, and P.R. LeClair, J.~Phys.: Condens.~Matter {\bf 15}, R109 (2003).


\bibitem{Miyazaki:55.1}
T.~Miyazaki and N.~Tezuka, 
J.~Magn.~Magn.~Mater.~{\bf 139}, L231 (1995);
J.S.~Moodera {\em et al.}, 
Phys. Rev. Lett. {\bf 74}, 3273 (1995).

\bibitem{Mathon:01.1}
J.~Mathon and A.~Umerski, Phys.~Rev.~B {\bf 63}, 220403(R) (2001);
W.H.~Butler, X.G.~Zhang, T.C.~Schulthess, and J.M.~MacLaren,
Phys.~Rev.~B {\bf 63}, 054416  (2001).



\bibitem{Yuasa:05.1}
S.~Yuasa {\em et al.}, 
Appl. Phys. Lett. {\bf 87}, 222508 (2005).


\bibitem{Zhang:05.1} 
X.G.~Zhang and W.H.~Butler,
Phys.~Rev.~B {\bf 70}, 172407  (2005).





\bibitem{Schneider06}
L.-N.~Tong {\it et al.}, Phys.~Rev.~B {\bf 73}, 214401  (2006).




\bibitem{Yuasa:06.1}
S.~Yuasa {\em et al.}, 
Appl. Phys. Lett. {\bf 89}, 042505 (2006).






\bibitem{Julliere:75.1}
M.~Julliere, Phys.~Lett.~A {\bf 54}, 225 (1975).

\bibitem{Mavropoulos00}
Ph.~Mavropoulos, N.~Papanikolaou, and P.H.~Dederichs,
Phys.~Rev.~Lett.~{\bf 85}, 1088 (2000);
D.~Wortmann, G.~Bihlmayer, and S.~Bl\"ugel,
J.~Phys.: Condens.~Matter {\bf 16}, S5819 (2004).


\bibitem{Itoh00}
H.~Itoh, T.~Ohsawa, and J.~Inoue, Phys.~Rev.~Lett.~{\bf 84}, 2501 (2000);
M.~Le\v{z}ai\'{c}, Ph.~Mavropoulos, J.~Enkovaara, G.~Bihlmayer and S.~Bl\"ugel,
Phys.~Rev.~Lett.~{\bf 97}, 026404 (2006).

\bibitem{Halilov98}
S. V. Halilov, H. Eschrig, A. Y. Perlov, and P. M. Oppeneer,
Phys.~Rev.~B {\bf 58}, 293 (1998).

\bibitem{Liechtenstein87}
A.I. Liechtenstein, M.I. Katsnelson, V.P. Antropov, and V.A. Gubanov,
J.~Magn.~Magn.~Mater.~{\bf 67} 65 (1987).






\bibitem{GGA}
J.P.~Perdew and Y.~Wang, Phys.~Rev.~B {\bf 45}, 13244 (1992).


\bibitem{SPRTBKKR} The SPR-TB-KKR package, H. Ebert and R. Zeller,
http: //olymp.cup.uni-muenchen.de/ak/ebert/SPR-TB-KKR

\bibitem{FLEUR}
Implemented in the {\tt FLEUR} code, http://www.flapw.de


\bibitem{footnoteb} \Tc\ cannot be measured for bcc
Fe$_{1-x}$Co$_x$ with high Co content, since these
alloys transform to fcc at about 1200~K [R.~Forrer, J.~Phys.~Radium
{\bf 1}, 49 (1930)]. Hcp Co also transforms to fcc near 700~K.\cite{Bozorth51}

\bibitem{Bozorth51}
R.M.~Bozorth, \textit{Ferromagnetism}, Van Nostrand, New York, 1951.

\bibitem{Ellis41}
W.C.~Ellis and E.S.~Greiner, Trans.~Am.~Soc.~Met.~{\bf 29}, 415 (1941);
G.A.~Prinz, Phys.~Rev.~Lett.~{\bf 54}, 1051 (1985).



\bibitem{Callaway}
D.~Bagayoko, A.~Ziegler, and J.~Callaway, Phys.~Rev.~B {\bf 27}, 7046 (1983).

\bibitem{footnotea}
Calculations show that the moment of bcc Co is also stable against non-collinear 
magnetic distortions [L.M.~Sandratskii and J.~K\"ubler, Phys.~Rev.~B {\bf 47}, 5854 (1993)].

\bibitem{Bardos69}
D. I. Bardos, J. Appl. Phys. {\bf 40}, 1371 (1969) 

\bibitem{Collins63}
M.F.~Collins and J.B.~Forsyth, Philos.~Mag.~{\bf 8}, 401 (1963).


\bibitem{KublerBook}
J.~K\"ubler, {\it Theory of itinerant electron magnetism}, Oxford
University Press (2000).

\bibitem{Diaz06}
A.~D\'iaz-Ortiz, R.~Drautz, M.~F\"ahnle, H.~Dosch~, and H.M.~Sanchez,
Phys.~Rev.~B {\bf 73}, 224208 (2006);
A.Y.~Liu and D.J.~Singh, Phys.~Rev.~B {\bf 46}, 11145 (1992);
K.~Schwarz and D.R.~Salahub, Phys.~Rev.~B {\bf 25}, 3427 (1982).




\bibitem{Lizarraga04}
R. Liz\'arraga, L. Nordstr\"om, L. Bergqvist, A. Bergman, E. Sj\"ostedt, P. Mohn, and O. Eriksson, Phys.~Rev.~Lett.~{\bf 93}, 107205 (2004).




\end{thebibliography}
\end{document}